\documentclass[fleqn,twoside]{article}
\usepackage{espcrc2}

\usepackage{graphicx}
\usepackage[figuresright]{rotating}

\usepackage{psfrag}
\usepackage{cite}

\newcommand{\AmS}{{\protect\the\textfont2
  A\kern-.1667em\lower.5ex\hbox{M}\kern-.125emS}}

\newcommand{\scsz}{\scriptsize}

\title{
       \begin{flushright}
	\vspace*{-12mm}
	{\normalsize UTCCS-P-10}\vspace{-3mm}\\ 
	{\normalsize UTHEP-498}\vspace{-3mm}\\
	{\normalsize KEK-CP-157}\vspace{-2mm}\\
       \end{flushright}
       Light hadron spectrum in 2+1 flavor full QCD
       by CP-PACS and JLQCD Collaborations
       \thanks{Presented by T.~Ishikawa}
       }

\author{CP-PACS and JLQCD Collaborations: 
        T.~Ishikawa\address[CCS]{Center for Computational Sciences, 
			         University of Tsukuba, Tsukuba, 
                                 Ibaraki 305-8577, Japan},
	S.~Aoki\address[Tsukuba]{Graduate School of Pure and Applied
                                 Sciences, University of Tsukuba,
	                         Tsukuba, Ibaraki 305-8571, Japan},
        M.~Fukugita\address[ICRR]{Institute for Cosmic Ray Research, 
                                  University of Tokyo, 
                                  Kashiwa 277-8582, Japan},
        S.~Hashimoto\address[KEK]{High Energy Accelerator Research
	                              Organization (KEK),
                                      Tsukuba, Ibaraki 305-0801, Japan},
        K-I.~Ishikawa\address[Hiroshima]{Department of Physics, Hiroshima 
                                         University, Higashi-Hiroshima,
                                         Hiroshima 739-8526, Japan},
	N.~Ishizuka$^{\rm a,b}$, 
	Y.~Iwasaki\addressmark[Tsukuba], 
	K.~Kanaya\addressmark[Tsukuba], 
	T.~Kaneko\addressmark[KEK],
	Y.~Kuramashi$^{\rm a,b}$,
	M.~Okawa\addressmark[Hiroshima],
	T.~Onogi\address[YITP]{Yukawa Institute for Theoretical Physics,
                              Kyoto University, Kyoto 606-8502, Japan},
	N.~Taniguchi$^{\rm a,b}$, 
	N.~Tsutsui\addressmark[KEK], 
	A.~Ukawa$^{\rm a,b}$, 
	T.~Yoshi\'e$^{\rm a,b}$ 
}

\begin{document}

\begin{abstract}
CP-PACS and JLQCD Collaborations are carrying out a joint project of the
2+1 flavor full QCD with the RG-improved gauge action and
the non-perturbatively ${\cal O}(a)$-improved Wilson quark action.
This simulation removes quenching effects of all three light quarks,
which is the last major uncertainty in lattice QCD.
In this report we present our results for the light meson spectrum and
quark masses on a $20^3\times40$ lattice at the lattice spacing $a\simeq0.10$
~fm.
\end{abstract}

\maketitle

\section{Introduction}

Three-flavor full QCD simulation is a current crucial task of lattice
QCD, as it would enable us, for the first time, to study hadron physics
based on principles of QCD with no approximations.
To achieve this, the CP-PACS and JLQCD Collaborations are carrying out
simulations with dynamical degenerate u, d quarks and a s quark with
different quark mass.

In our previous work presented at Lattice 2003, we made a exploratory
study at a lattice spacing $a\simeq0.10$~fm on a small lattice
$16^3\times32$ \cite{CPPACS-JLQCD-1}.
In this article we report the new results for the light meson spectrum and
quark masses on a larger lattice of $20^3\times40$ at $a\simeq0.10$ fm.
We note that the analysis procedures have been made more systematic over
these adopted at the time of Lattice 2004, and results are, albeit
qualitatively the same, numerically different.

\section{Strategy of the project}

We employ the RG-improved gauge action and the non-perturbatively
${\cal O}(a)$ improved Wilson quark action determined for this gauge action
\cite{CPPACS-JLQCD-2}, and plan to perform simulations at three lattice
spacings, $a\simeq0.07$ fm, $0.10$ fm, and $0.122$ fm, which are at even
intervals in $a^2$, with a fixed physical volume $\sim (2.0 \mbox{fm})^3$.
We take five u, d quark masses for chiral extrapolations in the range of
$m_{PS}/m_V\simeq0.78 - 0.62$, and two s quark masses around
$m_{PS}/m_V\simeq0.7$ for interpolations.
These parameters are chosen from our experience of simulations at
$a\simeq0.10$ fm on a $16^3\times32$ lattice.

An estimate of computational resources available to this project,
including the Earth Simulator, and performance of the polynomial hybrid
Monte Carlo program \cite{JLQCD-1,CPPACS-JLQCD-3} we use indicates that
we can accumulate at least $5000$ trajectories at $a\simeq 0.10$ fm and
$0.122$ fm and $2000$ trajectories at $a\simeq 0.07$ fm, within 
two to three years, of which we are in the second year.

\begin{table*}[t]
 \begin{center}
  \caption{$\kappa$ parameters, statistics and
  measured meson masses (in lattice unit).}
  \label{TAB:parameters}
  $\small\begin{array}{l l l | l l l l l l}\hline\hline
   \kappa_{ud} & \kappa_s & \mbox{traj.} & m_{PS,LL} & m_{PS,LS} &
    m_{PS,SS} & m_{V,LL} & m_{V,LS} & m_{V,SS} \\ \hline

   0.13580 & 0.13580 & 5000 & 0.5390(09) & -          &
   -          & 0.7024(17) & -          & -          \\

   0.13610 & 0.13580 & 6000 & 0.4943(07) & 0.5098(07) &
   0.5249(07) & 0.6656(19) & 0.6764(18) & 0.6871(18) \\

   0.13640 & 0.13580 & 7500 & 0.4460(06) & 0.4788(06) &
   0.5100(05) & 0.6189(19) & 0.6414(17) & 0.6635(15) \\

   0.13680 & 0.13580 & 8000 & 0.3786(08) & 0.4382(07) &
   0.4921(06) & 0.5642(27) & 0.6029(22) & 0.6410(18) \\

   0.13700 & 0.13580 & 8000 & 0.3384(09) & 0.4140(08) &
   0.4799(07) & 0.5300(19) & 0.5776(16) & 0.6243(14) \\ \hline

   0.13580 & 0.13640 & 5200 & 0.5254(09) & 0.4947(10) &
   0.4626(11) & 0.6853(14) & 0.6636(15) & 0.6416(17) \\

   0.13610 & 0.13640 & 8000 & 0.4804(07) & 0.4643(07) &
   0.4477(07) & 0.6451(14) & 0.6340(15) & 0.6229(16) \\

   0.13640 & 0.13640 & 9000 & 0.4324(05) & -          &
   -          & 0.6045(17) & -          & -          \\

   0.13680 & 0.13640 & 9000 & 0.3614(06) & 0.3867(06) &
   0.4107(06) & 0.5441(18) & 0.5598(16) & 0.5754(15) \\

   0.13700 & 0.13640 & 8000 & 0.3221(07) & 0.3625(07) &
   0.3994(06) & 0.5164(24) & 0.5396(21) & 0.5625(19) \\ \hline\hline
  \end{array}$
 \end{center}
 \vspace*{-3mm}
\end{table*}

In this project, our first interest is the meson spectrum and light
quark masses, because our physical volume $(2.0 \mbox{fm})^3$ is not
sufficiently large for baryons.

\section{Results at $a\simeq 0.10$ fm}

\subsection{Simulations and analyses}

Among three lattice spacings, calculations at $a\simeq 0.10$ fm have
been completed by simulations carried out at $\beta=1.9$ with
$c_{\mbox{\scsz SW}}=1.715$ on a $20^3\times40$ lattice.
In Tab. \ref{TAB:parameters} we show the hopping parameters $\kappa_{ud}$
and $\kappa_s$ for u, d (Light) quarks and s (Strange) quark.
Meson masses are calculated for Light-Light (LL), Light-Strange (LS) and
Strange-Strange (SS) combinations of valence quark masses,
and are given in the table.
They are determined by a single hyperbolic-cosine correlated fit with
$t_{\mbox{\scsz min}}=10$.
Masses are stable against a variation of $t_{\mbox{\scsz min}}$.

Chiral fits are made to LL, LS and SS masses simultaneously ignoring
correlations among these masses.
We adopt polynomial functions in quark masses up to quadratic terms.
$\chi^2/\mbox{d.o.f.}$ ($1.1$ for PS and $0.9$ for V) is acceptable.
We also test the fits using masses normalized by Sommer scale
and find that the meson spectrum and quark masses are stable against the
change.

\subsection{Light meson spectrum}

The physical point is determined from experimental values of 
$m_{\pi}, m_{\rho}, m_K$ (K-input) or $m_{\pi}, m_{\rho}, m_{\phi}$
($\phi$-input).
The inverse of the lattice spacing $a^{-1}=1.97(4)$ GeV is independent of
inputs for the s quark mass.

Light meson masses turn out to be
\begin{eqnarray}
\hspace*{+8mm}
 m_{K^{\ast}}=884.4(3.0) ~\mbox{MeV}&& \mbox{(K-input)},\nonumber\\ 
 m_{\phi}=998.5(5.9)     ~\mbox{MeV}&& \mbox{(K-input)},\nonumber\\ 
 m_K=519.4(6.2)          ~\mbox{MeV}&& \mbox{($\phi$-input)},\nonumber\\ 
 m_{K^{\ast}}=895.1(0.3) ~\mbox{MeV}&& \mbox{($\phi$-input)}.
\end{eqnarray} 

\begin{figure}[t]
 \begin{center}
  \psfragscanon
  \includegraphics*[scale=0.31]
  {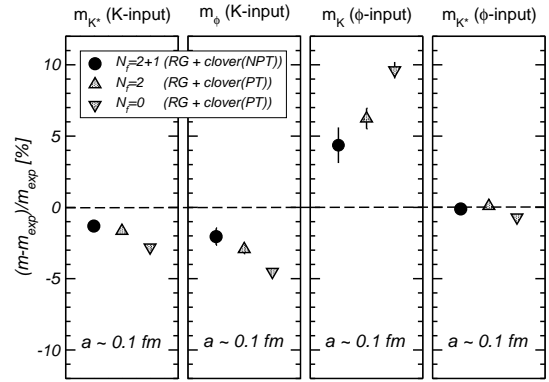}
 \end{center}
 \vspace*{-10mm}
 \caption{Deviation of the meson spectrum from experiment at
 $a\sim 0.1$ fm.}
 \label{FIG:Mmeson_vs_Nf}
 \vspace*{-5mm}
\end{figure}

In Fig.\ref{FIG:Mmeson_vs_Nf} we show the deviation of the meson
spectrum from experiment comparing with our results in $N_f=2$ and $0$ QCD
\cite{CPPACS-1} at the same lattice spacing $a\sim0.1$ fm.
We find that the spectrum in $N_f=2+1$ QCD is closer to experiment
than that in $N_f=2$ and 0 QCD at this lattice spacing.

\subsection{Quark masses}

Light quark masses are determined by the axial-vector Ward identity.
The matching to the $\overline{\mbox{MS}}$ scheme is performed by using
the mean-field improved one-loop calculation \cite{Aoki-1} at the scale
$\mu=a^{-1}$.
The renormalized quark masses are evolved to $\mu=2$ GeV using the 4-loop
beta function.

\begin{figure}[!h]
 \begin{center}
  \psfragscanon
  \includegraphics*[scale=0.30]
  {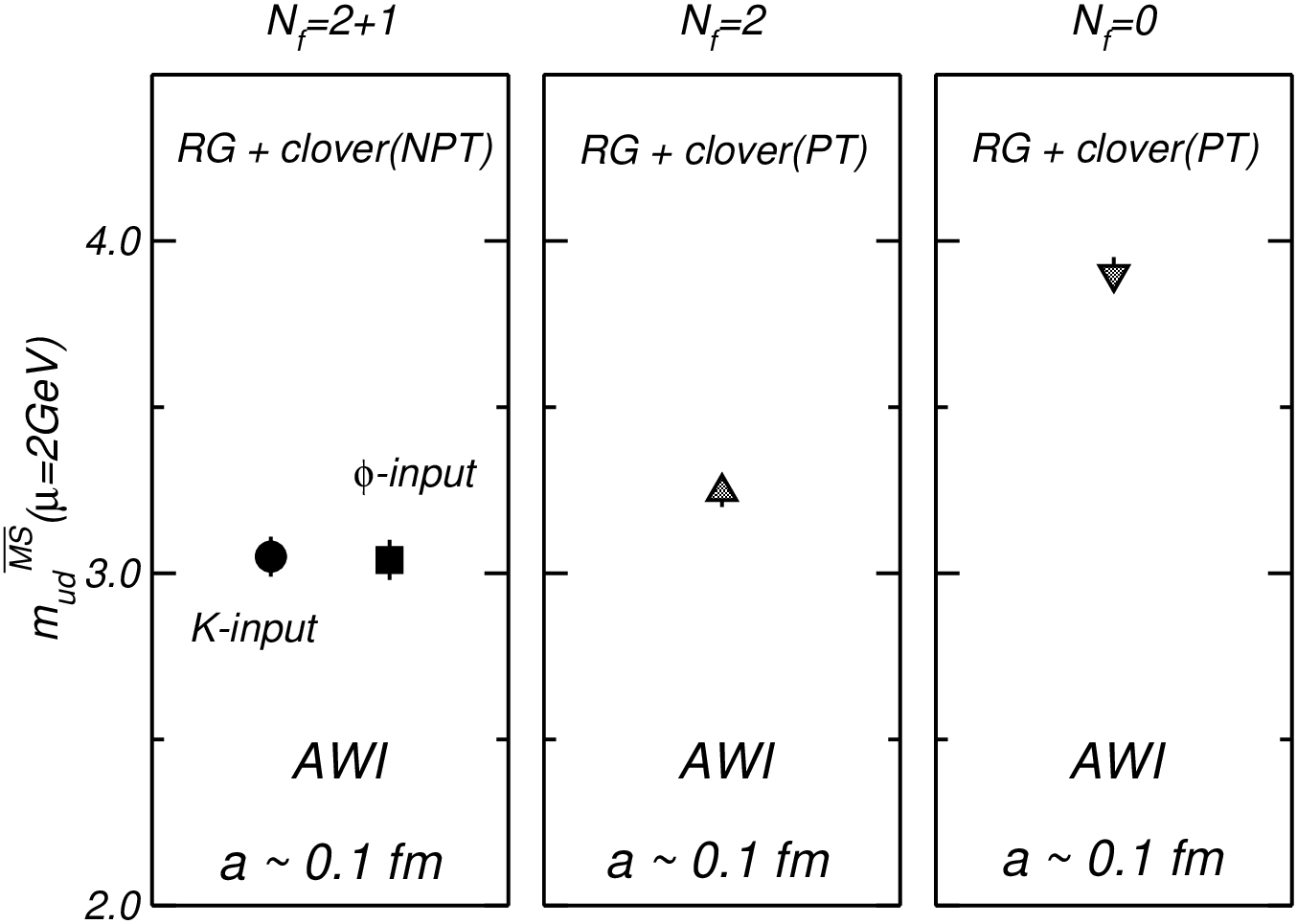}
 \end{center}
 \begin{center}
  \includegraphics*[scale=0.30]
  {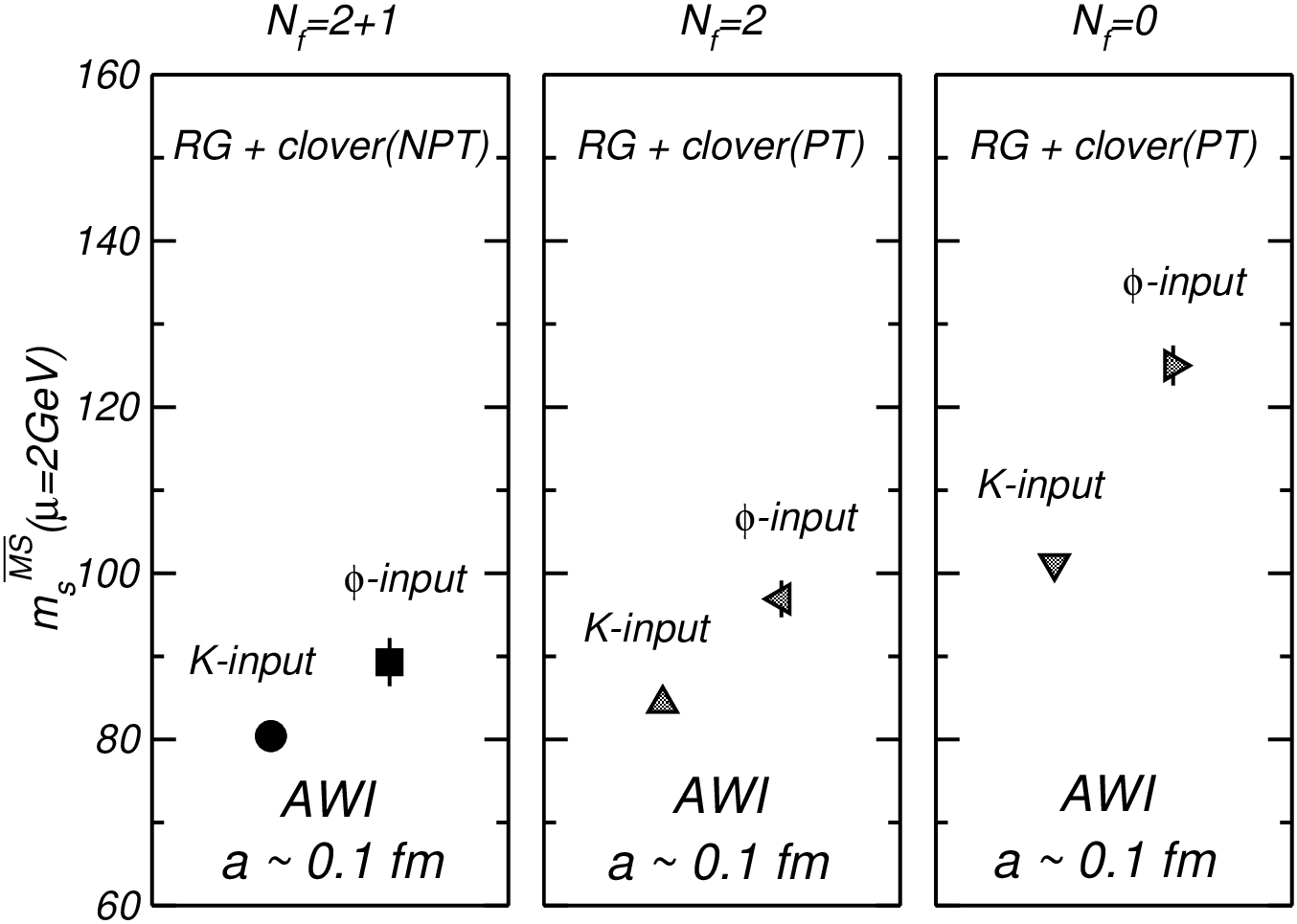}
 \end{center}
 \vspace*{-10mm}
 \caption{u and d quark masses (top) and s quark mass (bottom)
 at $a\sim0.1$~fm.}
 \label{FIG:Mq_vs_Nf}
 \vspace*{-3mm}
\end{figure}

Results for quark masses are
\begin{eqnarray}
 \hspace*{+9mm}
 \left\{\begin{array}{ll}
  m_{ud}=3.05(6)&\!\!\mbox{MeV}\\
  m_s=80.4(1.9) &\!\!\mbox{MeV}
       \end{array}\right.
 &&\hspace*{-5mm}\mbox{(K-input)},\nonumber\\
 \left\{\begin{array}{ll}
  m_{ud}=3.04(6)&\!\!\mbox{MeV}\\
  m_s=89.3(2.9) &\!\!\mbox{MeV}
       \end{array}\right.
 &&\hspace*{-5mm}\mbox{($\phi$-input)}.
\end{eqnarray}
Fig. \ref{FIG:Mq_vs_Nf} presents quark masses in $N_f=2+1$ QCD
comparing with $N_f=2$ and $0$ QCD at a similar lattice spacing
$a\sim0.1$ fm.
We find that quark masses in $N_f=2+1$ QCD are smaller than those in
$N_f=2$ and $0$ QCD. 
Since the non-perturbatively improved clover coefficient is used,
we speculate that the values of quark masses at this lattice spacing
do not change sizably in the continuum limit.
 
\section{Conclusions and future plans}

As far as we compare the spectrum and quark masses at $a\sim0.1$ fm,
we observe the difference between $N_f=2+1$ and $2$.
At this stage we cannot conclude whether this effect is due to dynamical
strange quark or non-perturbatively determined $c_{\mbox{\scsz SW}}$.
This point should become clear in the continuum limit.

Coupling values and hopping parameters for the coarser 
($a\simeq0.122$ fm) and finer ($a\simeq0.07$ fm) lattices have
already been fixed and production runs are in progress.
We hope to report results in the continuum limit in near future.\\ 

This work is supported by 
the Epoch Making Simulation Projects 
of Earth Simulator Center,
the Supercomputer Project No.98 (FY2003)
of High Energy Accelerator Research Organization (KEK),
the Large Scale Simulation Projects
of Academic Computing and Communications Center, University of Tsukuba,
Super Sinet Projects 
of National Institute of Infomatics,
and also by the Grant-in-Aid of the Ministry of Education
(Nos. 12740133, 13135204, 13640260, 14046202, 14740173, 15204015,
15540251, 15540279, 16028201, 16540228, 16740147).

\end{document}